# Wikinformetrics: Construction and description of an open Wikipedia knowledge graph dataset for informetric purposes

Wenceslao Arroyo-Machado[1*], Daniel Torres-Salinas[1], Rodrigo Costas[2,3]

[1] Department of Information and Communication Sciences, University of Granada, Granada, Spain

[2] Centre for Science and Technology Studies (CWTS), Leiden University, Leiden, The Netherlands

[3] DSI-NRF Centre of Excellence in Scientometrics and Science, Technology and Innovation Policy, Stellenbosch University, Stellenbosch, South Africa

* Corresponding author. Email: wences@ugr.es

## Abstract

Wikipedia is one of the most visited websites in the world and is also a frequent subject of scientific research. However, the analytical possibilities of Wikipedia information have not yet been analyzed considering at the same time both a large volume of pages and attributes. The main objective of this work is to offer a methodological framework and an open knowledge graph for the informetric large-scale study of Wikipedia. Features of Wikipedia pages are compared with those of scientific publications to highlight the (di)similarities between the two types of documents. Based on this comparison, different analytical possibilities that Wikipedia and its various data sources offer are explored, ultimately offering a set of metrics meant to study Wikipedia from different analytical dimensions. In parallel, a complete dedicated dataset of the English Wikipedia was built (and shared) following a relational model. Finally, a descriptive case study is carried out on the English Wikipedia dataset to illustrate the analytical potential of the knowledge graph and its metrics.

## Keyworkds

Wikipedia; Informetrics; Scientometrics; Altmetrics; metrics; indicators; knowledge graph; data; dataset

# 1. Introduction

On January 15, 2001 Wikipedia was born under the umbrella of Nupedia, an encyclopedia project whose edition was based on a peer review system. Due to the lack of agility in publishing articles, Wikipedia was created as a feeder project, as its objective was to make the creation of new articles easier before they were reviewed (*History of Wikipedia*, 2021). Wikipedia combined in a single project different elements that were new on the web and that made possible for the first time a universal encyclopedia (Reagle, 2009). It was successful enough to make Nupedia disappear in two years, experiencing a steady growth. Since then, Wikipedia has become one of the top visited websites of the world (https://www.semrush.com/website/top/, consulted on August 4, 2022), having 328 different editions, 285 of them having more than 1000 articles (https://meta.wikimedia.org/wiki/List_of_Wikipedias, consulted on August 4, 2022). Although this is the most successful project of Wikimedia Foundation, there are also other well-known knowledge projects using wikis as a basis (e.g., the Wiktionary dictionary or the Wikidata knowledge base).

Wikipedia has been a disruptive innovation, finding in its open nature and decentralized knowledge development one of its key elements (Olleros, 2008). Not only can everyone access its contents free of charge, but they can also participate in its construction, in a fully transparent process. This social construction of the knowledge can be seen in the differences found among language editions of the same Wikipedia pages (Hara & Doney, 2015). Wikipedia contents are also the result of consensus among editors or wikipedians. This consensus is built in open discussions in the so-called Wikipedia talks' pages (Maki et al., 2017; Yasseri et al., 2012), open to anyone and capturing transnational debates around Wikipedia contents (Kopf, 2020). Some of these talks and debates have sometimes transcended Wikipedia itself (O'Neil, 2017).

As an online encyclopedia, Wikipedia is not exempt from problems. The reliability of its content has been much debated since it is based on contributions from anonymous individuals (Olleros, 2008). The quality of Wikipedia pages' content has been studied numerous times from different perspectives, especially with regard to medical content pages, pointing out limitations such as occasional incomplete or imprecise information (C. E. Adams et al., 2020; Candelario et al., 2017; Weiner et al., 2019). The importance of integrating Wikipedia into academia, both in its use and in its development, has been highlighted (Jemielniak, 2019).

Social and cultural inequalities have also been pointed out, for example racial and gender gaps in its biographies (J. Adams et al., 2019; Tripodi, 2021).

Wikipedia is not free of bots and vandalism, although they do not constitute a serious threat to its contents and reliability and Wikipedia's policy does not allow detrimental use of the activity of bots or automated accounts. Most of the bots on Wikipedia are publicly identified (https://en.wikipedia.org/wiki/Special:ListUsers/bot), and they contribute to improving the content and structure of Wikipedia articles (Arroyo-Machado et al., 2020; Zheng et al., 2019). Bots also help to control and reduce problems of vandalism and trolls as they eliminate their harmful edits of articles in advance of human editors. There is also no shortage of proposals for methods based on machine learning to prevent this type of harmful activity (Martinez-Rico et al., 2019).

In spite of all previous issues, the general idea is that Wikipedia is a transparent and reliable source of encyclopedic information (Lageard & Paternotte, 2021), with value of its own to be subject of scientific research.

## 1.1. Wikipedia as source for informetric research

Wikipedia has been researched from different scientific perspectives. One of them is informetrics, quantitatively studying the contents and activity generated on Wikipedia. Thus, Wikipedia has been studied from the points of view of scientometrics, bibliometrics and webometrics, which are discussed in detail below.

Bibliographic references made in Wikipedia have been studied, particularly since the emergence of the notion of "altmetrics" (Priem et al., 2010), which considered citations on Wikipedia to scientific literature as part of its realm[1]. Wikipedia citations are one of the most popular sources covered in altmetric aggregators (Ortega, 2020; Zahedi & Costas, 2018) like Altmetric.com, PlumX or Crossref Event Data. In addition to altmetric data providers, there are also several other open data sources providing extensive metadata on Wikipedia citations (Singh et al., 2020; Zagorova et al., 2022). Moreover, other proposals like Scholia, enable exploring bibliographic data at different levels through Wikidata (F. Å. Nielsen et al., 2017). In Table 1 a summary of previous studies on Wikipedia bibliographic references are presented.

---

[1] Although Wikipedia references had been already studied for years before the birth of altmetrics, like the citation analysis by F. A. Nielsen (2007) or, in a more qualitative way, that of Mühlhauser and Oser (2008).

**Table 1.** Main studies on the bibliographic references included in Wikipedia pages.

| Reference | Type | Application | Data | Methodological approach | Language edition | Topic analized |
|---|---|---|---|---|---|---|
| *Mühlhauser and Oser* (Mühlhauser & Oser, 2008) | Qualitative | Content and quality analysis | --- | Check list | German | Health care |
| *Candelario et al.* (Candelario et al., 2017) | | Content and quality analysis | 33 pages | Scoring system | English | Medication |
| *Kaffee and Elsahar* (Kaffee & Elsahar, 2021) | | Analyze the editors' citation process | --- | Survey and interviews | Multilingual | Multidisciplinary |
| *Nielsen* (F. A. Nielsen, 2007) | Quantitative | Analyze citation patterns | 30,368 citations | Descriptive statistics | English | Multidisciplinary |
| *Kousha and Thelwall* (Kousha & Thelwall, 2017) | | Evaluate the impact of references | 36,191 citations | Descriptive statistics | Multilingual | Multidisciplinary |
| *Lewoniewski et al.* (Lewoniewski et al., 2017) | | References coverage across languages | 6.8 million pages 41 million citations | Descriptive statistics | Multilingual | Multidisciplinary |
| *Maggio et al.* (Maggio et al., 2017) | | Analyze citation patterns | 229,857 pages 1,049,025 citations | Descriptive statistics | English | Medicine |
| *Pooladian and Borrego* (Pooladian & Borrego, 2017) | | Evaluate the impact of references | 982 citations | Descriptive analysis | Multilingual | Multidisciplinary |
| *Jemielniak et al.* (Jemielniak et al., 2019) | | Rank journals by citations | 11,325 pages 137,889 citations | Citation analysis | English | Medicine |
| *Torres-Salinas et al.* (Torres-Salinas et al., 2019) | | Mapping of knowledge structure | 25,555 pages 41,655 citations | Co-citation analysis | English | Arts & Humanities |
| *Arroyo-Machado et al.* (Arroyo-Machado et al., 2020) | | Mapping of knowledge structure | 193,802 pages 847,512 citations | Co-citation analysis | English | Multidisciplinary |
| *Colavizza* (Colavizza, 2020) | | Publications coverage | 3,083 ref. pub. | Topic modeling and regression analysis | English | COVID-19 |
| *Nicholson et al.* (Nicholson et al., 2021) | | Reviewing citation quality | 1,923,575 pages 824,298 ref. pub. | Classification modeling | English | Multidisciplinary |
| *Singh et al.* (Singh et al., 2020) | | Dataset creation | 4 million citations | Text mining | English | Multidisciplinary |
| *Zagorova et al.* (Zagorova et al., 2022) | | Dataset creation | 6,073,708 pages 55 million citations | Text mining | English | Multidisciplinary |

Kaffee and Elsahar (2021) explored the flow that wikipedians follow to include references in Wikipedia articles. Kousha and Thelwall (2017), and Pooladian and Borrego (2017) described the problems of Wikipedia citations in performance evaluation. Nicholson et al. (2021) studied the quality of cited references in Wikipedia. Lewoniewski et al. (2017) showed that the different language editions of the same Wikipedia page tended to cite common sources, with the largest overlap between English and German; and some differences depending on the topics. Colavizza (2020) studied the coverage of the scientific literature on COVID-19 on Wikipedia, showing that although there was only a small percentage of scientific literature on COVID-19 in Wikipedia, it was sufficiently representative of its various topics. Arroyo-Machado et al. (2020) and Torres-Salinas et al. (2019) mapped Wikipedia co-citations patterns, showing fundamental differences in the use of scientific literature in Wikipedia compared to the academic realm. Bould et al. (2014), Li et al. (2021), and Tomaszewski and MacDonald (2016) studied academic citations in scientific publications to Wikipedia articles, proving that scientific publications also use Wikipedia content in their citations, as well as other digital encyclopedias, especially in areas such as Chemistry, Physics or Mathematics.

Wikipedia has also been the subject of webometric studies. For example, "*Wikiometrics*" were proposed as a rating system to rank universities or journals based on the features of their Wikipedia pages, also finding positive correlations with existing academic rankings (Katz & Rokach, 2017). The estimation of the importance of Wikipedia pages based on the PageRank algorithm was also studied, correlating positively with other page-view-based rankings (Thalhammer & Rettinger, 2016). Miquel-Ribé and Laniado (2018) showed that the different language editions of Wikipedia pages reflect cultural differences, as the contents cover local topics corresponding to different linguistic regions. Other studies focused on metrics about the attention generated around Wikipedia articles (e.g., likes or page view counts), showing how they reflect current topics of interest at a particular time/region (Dzogang et al., 2016; Mittermeier et al., 2019, 2021; Roll et al., 2016; Vilain et al., 2017), and even demonstrating the potential of Wikipedia pages to monitor the spread of diseases (Generous et al., 2014).

There are also numerous studies around Wikipedia's informetric features. Wilkinson and Huberman (2007) found a correlation between the quality of Wikipedia articles and their number of edits. The relationship between the length of Wikipedia articles and their quality has been highlighted by Blumenstock (2008). Beyond quality, relationships between Wikipedia metrics have also been explored. Previous studies found positive correlations between views and the number of edits and editors (Mittermeier et al., 2021), and weak correlations between the length of Wikipedia pages and the length of their talk pages (Yasseri et al., 2012). Zhang et al. (2018) suggested the value of using metrics in specific moments of the life cycles, for example the number of editors in the first three months of an article's life was not when it was most strongly related to its future quality.

Although as shown above there is abundant scientific literature on Wikipedia and its informetric applications, most of previous studies tended to focus on either limited sets of metrics (e.g. Nicholson et al. (2021) who were focused on the level of quality of scientific publications referenced in Wikipedia articles), or limited datasets (e.g. Mittermeier et al. (2021) who studied a large set of features in a dataset of Wikipedia pages of 10,099 bird species across 251 language editions). Thus, the large-scale study of Wikipedia, both from a large volume of pages and attributes, is still missing in the literature. Arguably, a potential reason for this lack of large-scale studies on Wikipedia is the lack of a conceptual framework that highlights both the large-scale data available from Wikipedia, as well as the multiple informetric metrics that Wikipedia offers. Such absence has hindered the development of broader research perspectives, especially regarding the relationship of Wikipedia with Science, where a contextualization of the relationships between the two is still needed.

In this study we propose such a framework by means of developing an informetric-inspired knowledge graph, with the aim of enabling similar analytical approaches as those developed in scientometric research. Such knowledge graph could work as complement of other Wikipedia knowledge graph like Wikidata (https://www.wikidata.org/) or DBpedia (https://www.dbpedia.org/). Wikidata and DBpedia provide exhaustive Wikipedia knowledge graphs but they are more focused on content and semantic relationships, transforming Wikipedia pages into entities (e.g., people, places, music bands, etc.) and establishing different computer-understandable relationships between them. Our proposed knowledge graph aims at characterizing the attention and usage of Wikipedia pages, using a relational model and incorporating activity metadata do not present in the semantic graphs of Wikidata and DBpedia, capturing the attention and social engagement, such as views or edits, as well as the presence of scientific literature in Wikipedia pages.

The paper is structured as follows: First, we describe our main objectives and our alignment with recent developments in the field of altmetrics. Second, we describe the informetric features of Wikipedia pages and their similarities with scientific publications, together with the existing data sources for data collection. Several informetric-inspired metrics (Wikinformetrics) are proposed for Wikipedia. Third, a Wikipedia knowledge graph, based on the combination of different Wikipedia data sources, is constructed and presented. Fourth, the dataset is explored in a descriptive way to show the analytical possibilities of the knowledge graph and the proposed metrics. Finally, we conclude by discussing our findings and proposing future research venues.

## 1.2. Objectives

The main objective of this work is to explore the research value of Wikipedia from an informetric perspective, and ultimately providing a complete Wikipedia knowledge graph. More specifically three objectives of different nature are targeted:

1. Theoretical objective: To establish a framework for Wikipedia analytics, by exploring the informetric features of Wikipedia pages (composition, categories, sources, data gathering, etc..) and proposing a set of informetric-inspired metrics (*Wikinformetrics*) for their quantitative study. This objective will help us mapping the analytical possibilities of Wikipedia as a scientific object.

2. Instrumental objective: To create a large open Wikipedia knowledge graph. Once we are familiar with the main features of Wikipedia, we will construct a dedicated knowledge graph focused on the English-language edition of Wikipedia with the main information and data relationships coming from combining different data sources.

3. Applied objective: To conduct a descriptive quantitative study of Wikipedia metrics based on the knowledge graph dataset, and to explore the proposed metrics and the different types of attention they capture.

This work and its objects align with novel developments on social media metrics (Díaz-Faes et al., 2019; Wouters et al., 2019), contributing to the exploration of different science-society interactions that can be captured on Wikipedia (Costas et al., 2020). Our ambition is to frame Wikipedia as a data source with multiple informetric research possibilities. Furthermore, a dedicated dataset of the English edition of Wikipedia is constructed for informetric purposes and is freely available at Zenodo (doi:10.5281/zenodo.6346899). R and Python were used together for its elaboration, with the scripts available on GitHub (doi:10.5281/zenodo.6959428). Many of the results presented here are novel, as to the best of our knowledge there is no previous literature that has explored the same large set of Wikipedia features and with the same large-scale perspective as in this study. This work is intended to be useful for a wide range of researchers, such as librarians, informetricians, sociologists or data scientists, among others.

## 2. Wikipedia from an informetric perspective

### 2.1. Analogy between Wikipedia pages and scientific publications

In Wikipedia the key component are the individual pages. Wikipedia pages are not only used for the publication of encyclopedia articles, but also other numerous typologies of pages, such as categories, users, talk pages, etc., as well as relationships among them. The different types of pages are given by a pre-established namespace (a type of page with special features identifiable through a prefix included in the title). Wikipedia currently has 12 namespaces in use (*article*, *user*, *Wikipedia*, *file*, *mediawiki*, *template*, *help*, *category*, *portal*, *draft*, *timedtext*, and *module*), each with an associated "talk namespace" (or "talk page") in which discussions are held around the contents and edits of the page, and 2 virtual namespaces (special and media).

There are several features of Wikipedia pages, in particular namespace article pages, for which it is possible to establish an equivalence with that of a scientific publication. First, they have a title and an associated page identifier (Wikipedia page id). They may have one or more authors, being possible to identify the first who created it, and when, and those who have made a greater contribution or whose edition has been revoked. The contents may include multimedia files, links to external resources, and bibliographic references, among others. There are also internal links that enable connecting Wikipedia pages to each other, just like citations among scientific publications. Finally, Wikipedia pages can be classified with categories according to their contents to carry out its thematic classification, like keywords and classifications applied to scientific publications. Most of these elements can be seen as metadata to be treated in the study of Wikipedia pages. However, there are several differences between Wikipedia pages and scientific publications that cannot be ignored (Table 2). The most important is that Wikipedia pages are a living resource and not a static document. The access and editing of the contents also differ between Wikipedia pages and scientific publications, since Wikipedia pages do not focus on a specific audience (e.g., scientific publications mostly focus on academic audiences), but anyone can take an active part in editing them. It should be also noted that some pages may be temporarily limited or protected for editing (Hill & Shaw, 2015).

**Table 2.** Comparison of features between Wikipedia pages and scientific publications.

| Wikipedia element description | | Wikipedia pages vs. Scientific publications | |
|---|---|---|---|
| | | Wikipedia page | Scientific publication |
| *State* | Document state condition | Living | Static |
| *ID* | Document identification number | Page ID | DOI, ISBN, URI… |
| *Name* | Title of the document | Title | Title |
| *Type* | Document typologies | Namespace (12+12 types) | Paper, proceeding, letter… |
| *Creation* | Date from which it is available | First edition date | Publication date |
| *Authorship* | Responsables of the work | Wikipedians | Authors |
| *Content* | Type of content | Structured text | Structured text |
| *Language* | Language of the resource | Edition dependent | Document dependent |
| *Discussion* | Comments on the contents | Talk | Peer review |
| *Description* | Work summary | Short description | Abstract |
| *Tags* | Terms describing the content | Categories | Keywords |
| *Media* | Audiovisual resources includible | Images, audios, and videos | Images, audios, and videos |
| *Internal links* | Links to the related resources | Internal links | Citations |
| *Format* | Standardized structure and content | Manual of style[*] | Format guidelines |
| *Bibliography* | References of cited resources | References | References |
| *Access* | Access model | Open | Closed/Open |
| *Audience* | Document target audience | General | Specialized |
| *The English Wikipedia has its own manual of style https://en.wikipedia.org/wiki/Wikipedia:Manual_of_Style | | | |

The living nature of Wikipedia pages puts them at the center of a complex system (Ladyman et al., 2013), whose main elements are represented in Fig 1. Many of the elements of the pages are static or unalterable, such as the creation date or page id, while others are in constant evolution, especially the contents themselves. This makes it difficult to study certain elements in Wikipedia (Détienne et al., 2016), since Wikipedia content is volatile, and authorship and contribution roles can be diluted in contrast to the higher stability of scientific publications. In addition, the same page, especially encyclopedic articles, may have parallel versions in different language editions of Wikipedia, which may vary in content. This scenario becomes even more complex when taking into account that not only human users are involved in the development of Wikipedia pages, but also bots, thus making the interactions that can occur more complex to analyze (Tsvetkova et al., 2017).

**Fig 1.** Diagram of the main elements involved in creating and editing Wikipedia articles.

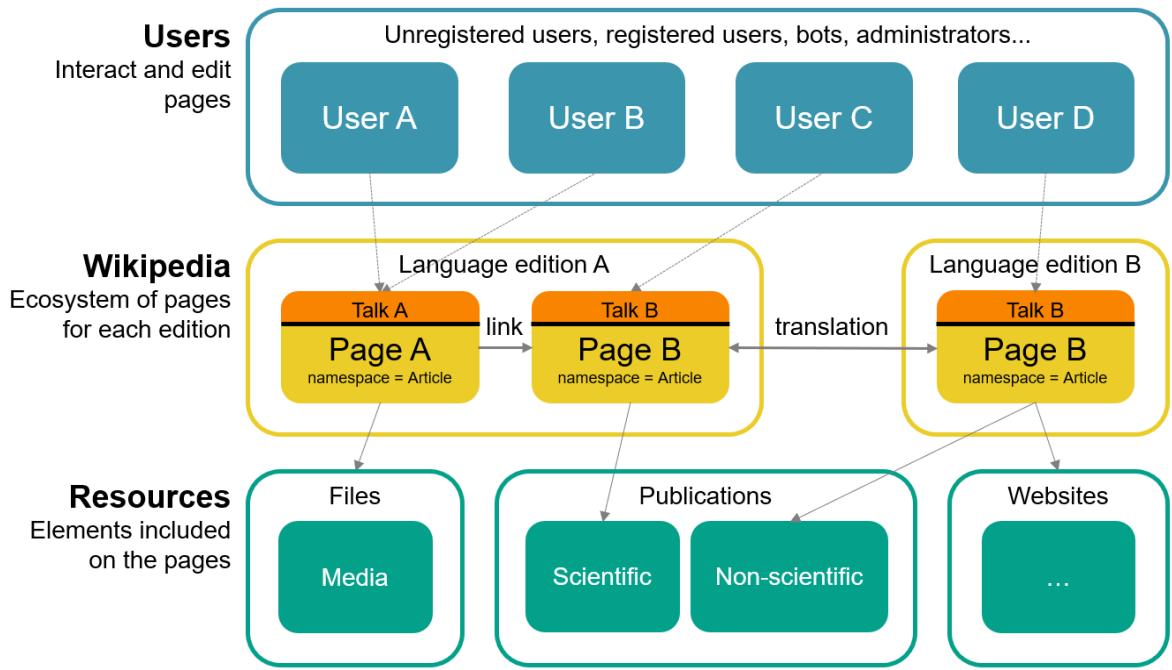

## 2.2. Categorization

Wikipedia pages are not thematically organized according to a controlled language-based classification, such as Britannica's subject organization system. Instead, Wikipedia pages have a category system that works like a folksonomy (Minguillón et al., 2017). Wikipedians are free to tag each page under one or more existing categories or to create new ones. Numerous studies have approached them, for example, by studying their semantic domain (Aghaebrahimian et al., 2020; Heist & Paulheim, 2019). However, the main problem of this folksonomy is the large number of individual categories and their unstructured (i.e. without a clear hierarchical system) relations at different levels, introducing a lot of noise and making it difficult to have a general thematic view of Wikipedia (Boldi & Monti, 2016; Kittur et al., 2009). In addition, there are also hidden categories, related to the maintenance or management of the page.

Besides the categories, Wikipedia has other options for accessing and browsing its contents by topics (https://en.wikipedia.org/wiki/Wikipedia:Contents). On the one hand, it offers different curated content lists (e.g., the "list of articles every Wikipedia should have" or the list of "vital articles"). There are other lists that offer collections of articles that respond to the same topic, and even "lists of lists". Similarly, there are "portals", which imitate the classic web portals and are organized in sections that group the main contents of a topic, not only the articles (e.g., the "Science" portal or the "History of science" subportal). WikiProjects, communities of wikipedians aimed at improving Wikipedia content on a specific topic and which have their own page from which they coordinate their activities, can also work as a classification approach due to their thematic orientation (e.g., "Anthropology" or "The Beatles"). There are also third-party classification systems, such as the "Library of Congress Classification" or the "Universal Decimal Classification". Finally, external to Wikipedia, but within the Wikimedia ecosystem, there are other types of classification solutions, such as Wikidata taxonomies ([https://www.wikidata.org/wiki/Wikidata:WikiProject_Taxonomy](https://www.wikidata.org/wiki/Wikidata:WikiProject_Taxonomy)) or ORES ([https://www.mediawiki.org/wiki/ORES](https://www.mediawiki.org/wiki/ORES)), that can be used to identify Wikipedia pages topics using machine learning techniques. The main limitation with all of the above, is that there is no central classification system that covers all Wikipedia pages, and that at the same time it is concise and easy to manage, particularly in terms of the number of subjects and the hierarchical relationships among them. The lack of such central classification in Wikipedia is a major hindrance for the large-scale epistemic study of Wikipedia.

## 2.3. Content-control

Each Wikipedia page has a discussion space called "talk pages", where wikipedians discuss with other wikipedians. Talk pages aim at improving the quality and reliability of the articles. Discussions in talk pages are public (Ferschke et al., 2012), resembling the model of open peer review of scientific publications (Black, 2008), and representing a form of public review in contrast to the traditional academic blind peer review system (Cummings, 2020). Wikipedia also counts with formal peer review approaches in which wikipedians request assistance from experts on given topics ([https://en.wikipedia.org/wiki/Wikipedia:Peer_review](https://en.wikipedia.org/wiki/Wikipedia:Peer_review)). Despite discrepancies and differences about what open peer review means and the different models proposed (Ross-Hellauer, 2017), the three basic principles (open identities, reports, and participations) are clearly recognizable in Wikipedia (S2 Table). Wikipedians are both authors and reviewers of content and their reports are available as comments on the talk pages, all of which are always open and identifiable. Interestingly, Wikipedia-inspired reviewing approaches have even been proposed for scholarly publishing, such as the post-publication correction system and readers' comments (Xiao & Askin, 2014).

Wikipedia also counts with a quality control system of the content of the different articles that comes from WikiProjects. It is grounded on an evaluation system to classify pages in higher or lower levels of content quality, with standard grades, which are listed on the respective talk page. Although there is a general scheme (Table 3), it is possible that some WikiProjects do not include all grades or that there may be differences in their application. Similarly, the pages are also classified according to their importance within the topic (Top, High, Mid, and Low). Wikipedians can set any level of quality and importance on a given page, as well as to modify them. When there are disagreements among wikipedians in the quality levels of a page, this leads to a discussion and quest of consensus around the quality level of the page. However, at the highest levels of quality (Featured Articles and Good Articles) this assignment requires a stricter review process, including the presentation of a candidature and an evaluation by independent wikipedians according to pre-established criteria. These two levels also have their own badges on the article page.

**Table 3.** General quality grading scheme of WikiProject articles.

| Class | Description | Assignment | Badge |
|---|---|---|---|
| **Featured article** | The best possible content on Wikipedia, no need for improvement | Review | Yes |
| **Featured list** | The best possible list on Wikipedia, no need for improvement | Review | Yes |
| **A** | Fully addresses the subject and requires only minor improvements | Review | No |
| **Good article** | It satisfies Wikipedia's main criteria and is close to a professional article | Review | Yes |
| **B** | The content is almost complete and has no major problems | Free | No |
| **C** | The content is considerable, but has significant problems | Free | No |
| **Start** | It includes significant content, but is still in development | Free | No |
| **Stub** | The content is very short and requires substantial work | Free | No |
| **List** | Content displayed in a list linking to Wikipedia articles on a specific topic | Free | No |

## 2.4. Sources

A fundamental aspect of Wikipedia lies in the system of links that allows its pages to be connected among them, making Wikipedia unique in this sense with regards to other encyclopedic systems (Reagle & Koerner, 2020). These internal links have been studied previously, showing both the semantic relationships they can establish and other potential utilities (Consonni et al., 2019; Presutti et al., 2014), as well as the possibility of calculating network indicators like PageRank based on them (Thalhammer & Rettinger, 2016). There are however important issues to consider when working with Wikipedia pages links:

1) The links may be redirects, i.e., old page versions that automatically redirect to the new versions when accessing them.

2) There are lists of links to other Wikipedia pages. Most of the lists include pages that are conceptually related to each other and share a clear subject matter, however there are specific lists such as disambiguation pages, which are aimed at reducing the ambiguity of some terms (e.g., "citation" or "granada"), and therefore the links in these lists are not necessarily thematically related.

Another fundamental source for Wikipedia is its bibliographic references. Wikipedia recommends the use of bibliographic references to support its contents and it is an essential requirement for a page to achieve the best quality status (Featured article). These references are the same as those made in scientific publications, in both cases serving as a support for an idea. However, it is necessary to consider that citations in Wikipedia and citations in scientific publications are governed by different norms and dynamics. In Fig 2 the main differences between scientific publications references and Wikipedia references are schematized.

**Fig 2.** Differences between traditional citations and Wikipedia mentions to scientific publications.

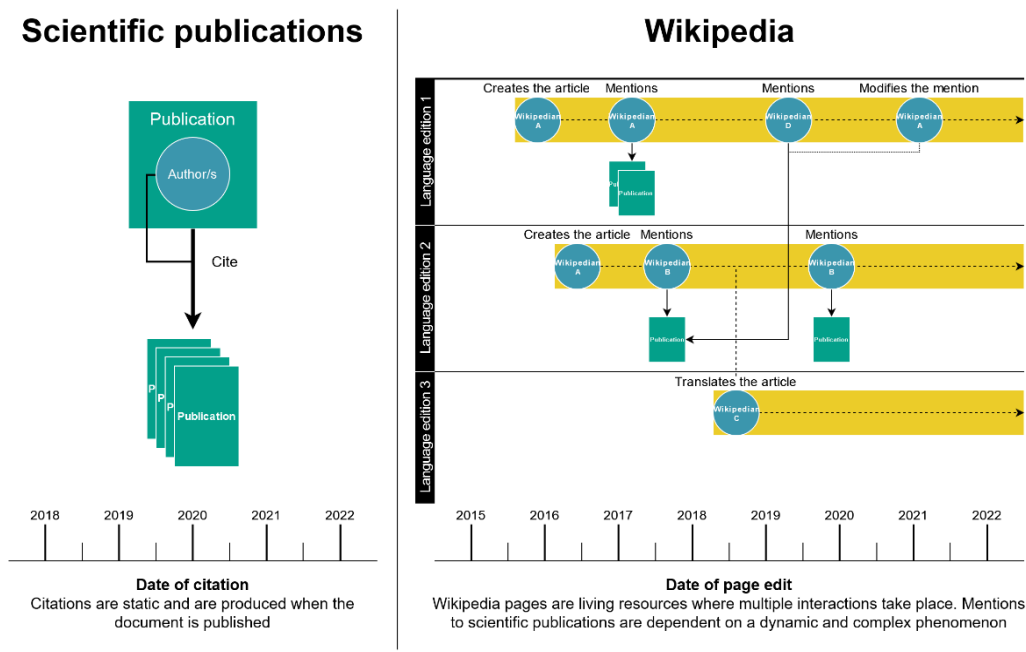

Other relevant particularities of Wikipedia references include:
- Unlike scientific publications in which the identity of the citers (i.e., those including the references in the scientific publication) is clear and invariable, in Wikipedia this is more complex (given the live nature of Wikipedia articles) and not always possible.

Although, there are some methodological proposals for this purpose (Zagorova et al., 2022).

- Wikipedia citation counts can be distorted by the translations of articles into different languages, since it is possible to easily transfer the references across the different language versions of the same article, thus distorting the meaning and value of Wikipedia citation counts. Such limitation does not occur in scientific publications, since only one language version of a given publication is usually considered in the counting of citations.
- There are certain Wikipedia pages that function as large bibliographic indexes, bringing together the most relevant literature on a specific topic (e.g., research annuals or bibliographies).
- There are also templates (special Wikipedia pages that are embedded within other pages to facilitate the repetition of information), which are sometimes used to generate pre-established lists of references that are quickly inserted and replicated into numerous Wikipedia pages that are strongly related. This happened for example with the listing of lunar crater references (https://en.wikipedia.org/wiki/Wikipedia:Templates_for_discussion/Log/2014_June_8#Template:Lunar_crater_references).

## 2.5. Data gathering

There are numerous data sources and the choice of one or the other depends mostly on the type and volume of data required. In some cases, there are even multiple ways of accessing the same data. These have been summarized in Table 4, but can be found in detail in S3 Appendix. In fact, Wikimedia has a Research community (https://meta.wikimedia.org/wiki/Research) that gathers different resources to help and guide all those people who want to access the data of the Wikimedia projects and that lists the different projects related to it.

**Table 4.** Summary of Wikipedia data sources by format, update frequency, data quantity, type, and challenges.

| | Content | Access | Format | Update frequency | Data quantity* | Type** | Main challenge*** |
|---|---|---|---|---|---|---|---|
| **Wikimedia Dumps** | Metadata, page content, and relationships | Offline | XML, SQL | Once/twice a month | Big data | General | Data processing |
| **MediaWiki and Wikimedia APIs** | Metadata, page content, relationships, and statistics | Online | JSON, WDDX, XML, YAML, PHP | Realtime | Small data | General | Data recovery |
| **Wiki Replicas** | Metadata, page content, and relationships | Online | SQL | Near-realtime | Small data | General | Data recovery |
| **Event Streams** | Real-time logs | Online | SSE, JSON | Realtime | - | Specific | Data recovery |
| **Analytics dumps** | Statistics on page views and activity | Offline | TSV | Monthly | Big data | Specific | Data processing |
| **WikiStats** | Statistics on page views, content and activity | Online | JSON/CSV | Monthly | Small data | Specific | Data recovery |
| **Dbpedia** | Contents and semantic relationships | Both | RDF/XML, Turtle, N-Triplets, SPARQL endpoint | Live/monthly | - | General | Data recovery |
| **XTools** | Statistics on page views, content and activity | Online | JSON | Realtime | Small data | Specific | Data recovery |
| **Repositories** | Dedicated Wikipedia datasets | Offline | - | - | - | - | - |
| **Altmetric aggregators** | Wikipedia References to publications | Online | CSV/JSON | Daily | - | Specific | Data processing |

*Volume of data to be retrieved and processed.

**Data from Wikipedia are included to address different problems or are of a specific nature.

***Task that will require more effort when using the data source.

The two main sources are dumps and APIs. One of the main problems when working with Wikipedia data dumps is their size, especially when dealing with the more complete editions (e.g., the metadata of the revision of the English Wikipedia pages as of June 2022 is formed by 27 files of more than 2GB each), so accessing a subset of data requires a lot of time and effort. In the case of using Wikipedia APIs, metadata can be accessed on demand, but the retrieval process is very laborious, especially when large volumes of data are required. Other sources are characterized by offering already preprocessed data, such as the total number of edits or page views, which can be consulted from XTool.

In this paper we extracted and developed a full Wikipedia knowledge graph with the ambition of facilitating the future of the English Wikipedia, reducing the time and effort that researchers may need in collecting and connecting all the different data sources.

## 2.6. *Wikinformetrics*

Finally, there are multiple metrics that can be extracted from the sources presented before and that enable the informetric study of Wikipedia pages. Based on previous studies and the above exploration of the informetric characteristics of Wikipedia, several metrics have been selected (Table 5). Each of them is of interest for measuring a particular dimension of the pages. For example, the number of views can be seen as a measure of the impact and outreach of a particular page, and while the number of edits and editors reflect the volume of activity, the number of talks and talkers are representative of the discussions that take place around these pages. These are not the only metrics that can be obtained from Wikipedia, but they can be considered to capture some of the most important analytical aspects of Wikipedia pages (e.g., contributions, content development, links and interactions, and impact), being also easy to interpret in an informetric framework.

**Table 5.** Description of the metrics obtained for Wikipedia articles by analytical dimension.

| Metric | Analytical dimension | Description |
|---|---|---|
| **Editors** | Activity | Number of unique editors that have edited a Wikipedia article |
| **Edits** | Activity | Number of total edits that have a Wikipedia article |
| **Linked** | Connectivity | Number of Wikipedia articles in which the article is linked to |
| **Links** | Connectivity | Number of internal links that include a Wikipedia article to others |
| **Age** | Description | Years that have passed since the creation of the page to the date of data collection |
| **Length** | Description | Length in bytes of the page |
| **Talkers** | Discussion | Number of unique editors that have edited a Wikipedia article's talk page |
| **Talks** | Discussion | Number of total edits that the talk page of a Wikipedia article has |
| **Views** | Outreach | Number of daily views of a Wikipedia page |
| **References** | Support | Number of elements listed in the references |
| **Pub. referenced** | Support | Number of publications referenced |
| **URLs** | Support | Number of external links that include a Wikipedia article |

## 3. Wikipedia knowledge graph

Using the different data sources described above, a knowledge graph of the English edition of Wikipedia has been constructed for informetric purposes and freely shared on Zenodo (doi:10.5281/zenodo.6346899). The English edition of Wikipedia has been chosen because it is the largest one and has an international scope. For its construction, data from Wikimedia and analytic dumps were used, as well as data shared in repositories, specifically the dataset of Singh et al. (2020) in which they share references made in Wikipedia articles. The data included in this dataset covers all English Wikipedia activity until July 2021, except page views, which are from April 1, 2021 to June 30, 2021, and bibliographic reference data, until May 2020. R and Python have been used together, with the scripts available on GitHub (doi:10.5281/zenodo.6959428). The construction of this dataset is described in S1 Appendix. The resulting dataset consists of 9 files connected to each other by a relational structure summarized in Fig 3.

**Fig 3.** Diagram of files and relationships of the Wikipedia knowledge graph dataset.

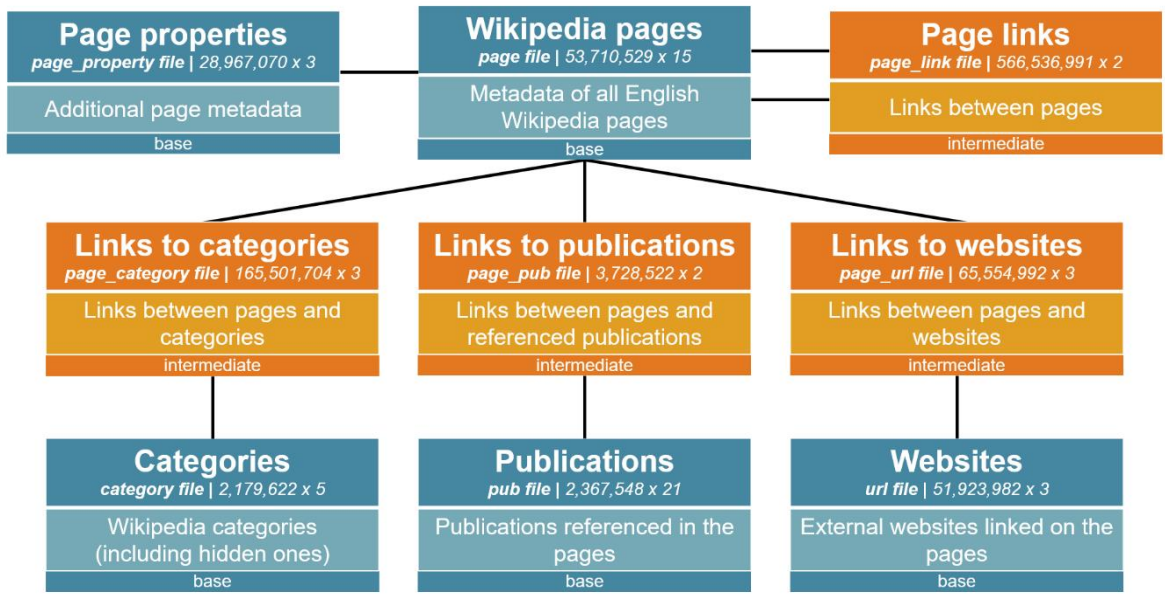

**Wikipedia Knowledge Graph**, dataset and description free at: **10.5281/zenodo.6346899**

This knowledge graph offers numerous possibilities for the informetric study of Wikipedia, making it possible to study new relationships (and interactions) between science and this social media (e.g., the attention on Wikipedia to academic topics, the presence of scientific literature on popular Wikipedia pages, or the use of scientific literature in Wikipedia pages with large discussions in their Talk pages, to name a few). This is the case of the work of Arroyo-Machado et al. (2022), who found a positive relationship between the research performance of universities and their social attention on Wikipedia, using data from this dataset.

Although the generation of new versions of the knowledge graph cannot be guaranteed by the authors of this paper, the way in which its creation is detailed, and the shared scripts ensure that new versions can be generated. This is also of importance for the generation of new knowledge graphs in other language editions of Wikipedia, as the data used as a basis is also available in other languages. The only limitation in this respect is in the reference data, as they come from a specific dataset (Singh et al., 2020). However, those responsible have also shared the tools used to obtain the references and there are other alternatives such as Zagorova et al. (2022) or altmetric data aggregators.

# 4. Case study: informetric analysis of the English Wikipedia

As a case study, the knowledge graph of the English Wikipedia is used to calculate and study the proposed metrics in a broad manner. The analysis was performed in Python and the code is available at GitHub (doi:10.5281/zenodo.6958972).

## 4.1. Wikipedia metrics and article's content

There are a total of 53,710,529 pages in the English Wikipedia, considering all namespaces as well as pages that are redirects, however this number is reduced to 6,328,134 pages when the focus is on articles that are not redirects. These represent just 11.79% of the overall English Wikipedia. For all of them, the metrics proposed in Fig 4 have been obtained.

**Fig 4.** Average of Wikipedia article metrics differentiating by the quality assigned from a project.

| N. of articles →<br>Wiki Metrics ↓ | All articles | Featured articles | Featured lists | A | Good | B | C | List | Start | Stub |
|---|---|---|---|---|---|---|---|---|---|---|
| | 6,328,134 | 5945 | 3816 | 958 | 34,004 | 109,019 | 394,065 | 253,066 | 1,818,356 | 3,079,778 |
| Editors | 48.38 | 516.93 | 179.13 | 176.80 | 275.71 | 297.62 | 165.36 | 56.27 | 63.13 | 22.85 |
| Edits | 101.92 | 1491.35 | 593.61 | 564.91 | 724.13 | 705.41 | 369.89 | 159.80 | 129.52 | 40.23 |
| Linked | 80.53 | 725.25 | 175.84 | 202.01 | 330.18 | 417.00 | 234.08 | 107.34 | 93.03 | 55.70 |
| Links | 87.77 | 329.68 | 270.16 | 236.56 | 224.88 | 233.87 | 164.23 | 174.78 | 101.28 | 69.90 |
| Age | 9.59 | 14.33 | 11.52 | 12.74 | 12.06 | 12.47 | 10.92 | 9.13 | 10.45 | 9.20 |
| Length | 7844.68 | 61,248 | 51,549 | 43,329 | 39,444 | 35,009 | 21,676 | 18,202 | 10,033 | 3748 |
| Talkers | 5.38 | 66.17 | 16.62 | 27.90 | 29.64 | 28.16 | 15.03 | 4.98 | 6.56 | 3.64 |
| Talks | 9.19 | 258.40 | 42.36 | 92.21 | 88.56 | 88.35 | 35.32 | 9.07 | 9.69 | 4.32 |
| Views | 3345.07 | 64,801 | 26,685 | 16,011 | 29,229 | 30,359 | 15,829 | 3777 | 4094 | 710 |
| References | 4.6 | 53.95 | 55.49 | 31.76 | 38.87 | 26.51 | 15.40 | 9.20 | 5.79 | 1.84 |
| Pub. Ref. | 0.59 | 14.27 | 2.34 | 8.51 | 5.83 | 4.77 | 2.37 | 0.53 | 0.69 | 0.22 |
| URLs | 10.33 | 58.03 | 67.32 | 33.32 | 46.10 | 40.31 | 25.95 | 22.82 | 12.90 | 6.09 |

Fig 4 shows the descriptive statistics of the main variables, differentiating between total Wikipedia articles and those classified based on their quality. A total of 5,522,676 articles (87.27% of the total) are associated with a WikiProject and with some quality level. Articles with different quality levels have been considered in all of them. It is noticeable that in all metrics Featured Articles have the highest values. The case of class B articles is noteworthy, as they not only show few differences with respect to the Good and A-Class articles, being also greater in number of articles than both, but in aspects such as views, they are positioned above them.

There are important differences in the number of referenced publications, going from an average of 14.27 publications in Featured articles to 8.52 in A and 5.84 in Good articles, while the Start and Stub articles cite on average less than one publication. This reflects compliance with English Wikipedia's criteria for establishing the quality level of articles. The general criteria do not make explicit the need for a greater number of references to increase the level of quality, among others, but they do require an increase in "reliable sources", so that citations to publications can serve as a proxy for this. Likewise, it also corroborates previous findings of a relationship between the level of quality and the number of edits (Wilkinson & Huberman, 2007), and the length of articles (Blumenstock, 2008).

Most of Wikipedia pages are not of recent creation (Fig 5A), with a median of 11 years. In some of the metrics, such as edits and talks, extreme outliers are found. This can be seen in the fact that their average values are 102 and 9.19, respectively, above the median and third quartile values. This situation is much more pronounced in the case of views, with an average of 3346.59. Furthermore, the number of referenced elements has a median of 1 and an average of 4.6. When comparing the links with the linked ones, we find that Wikipedia pages link more than they are linked, since the median for the former is 36 and for the latter 15.

**Fig 5.** A) Boxplots of the main metrics for Wikipedia articles excluding outliers from the figures and marking the mean with a cross symbol. B) Spearman's Rho correlations between the main metrics for Wikipedia English articles.

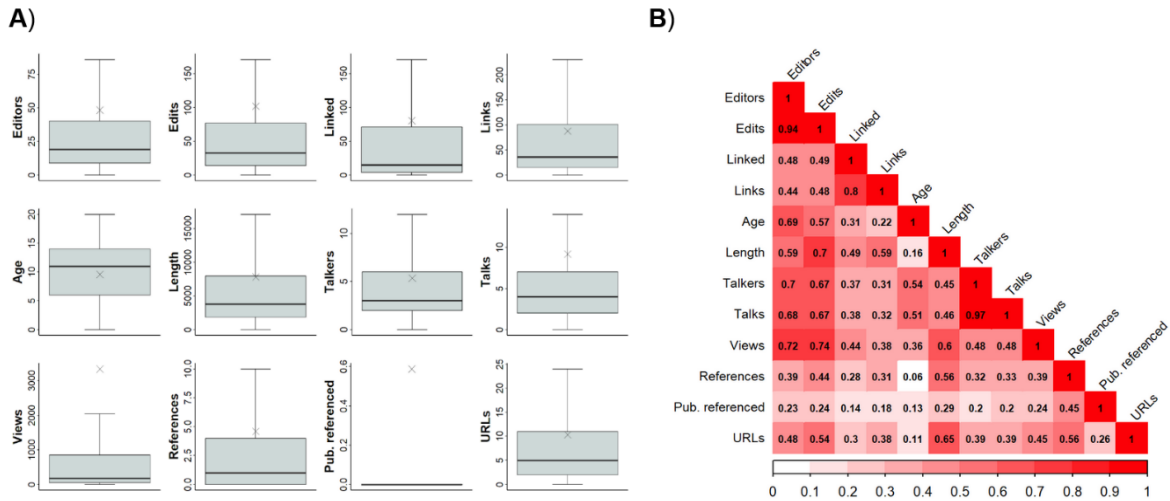

The correlations between these variables are all positive (Fig 5B). The strongest correlation is between talkers and talks ($r_s$=0.97), followed by another analogous relationship such as that between editors and edits ($r_s$=0.94). When considering pairs of metrics of different nature, the strongest correlation is between edits and views ($r_s$=0.74), followed by that of editors and views ($r_s$=0.72), which suggests a relationship between the popularity of Wikipedia pages in terms of visits and their number of edits. Interestingly, a lower correlation was found between views, and both talks and talkers ($r_s$=0.48), suggesting that discussions around Wikipedia pages are not necessarily related to higher number of views. Other moderate correlation can be found between the length of an article and its views ($r_s$=0.6), which may indicate that the larger the article the more attention it receives or that the more attention it receives the more it grows in length. There are other moderate correlations, such as between the length and the number of references ($r_s$=0.56) and URLs ($r_s$=0.65), but which are to be expected as the two elements directly interfere with each other. The number of referenced publications is the metric most weakly correlated, there being for example a weak correlation between this and views ($r_s$=0.24) or talks ($r_s$=0.2). Our results confirm the same type of relationships reported in previous research (Mittermeier et al., 2021), albeit this time considering the entire population of English language Wikipedia articles.

## 4.2. Different types of attention captured on Wikipedia

The results of this analysis can also be accessed interactively and in greater detail via R Shinny app: https://wenceslao-arroyo-machado.shinyapps.io/wikinformetrics/

A review of Wikipedia's main pages based on different metrics reveals its potential to capture content that responds to different types of attention (S4 Table). The page views make it possible to identify those topics that capture the most attention of society in a given period— page views are limited to a period of 3 months in our dataset—. Thus, in our dataset the pages of *Prince Philip, Duke of Edinburgh* (10,860,553 views) and *Elizabeth II* (9,900,275), or *Mare of Easttown* (5,995,513) rank among the most visited in the English language Wikipedia. Also, five of the twenty most viewed pages are series or movies released in the period analyzed, which also highlights that contents related to entertainment occupy a relevant position in Wikipedia. Sports also receive many views and reflect current events, as evidenced by the *UEFA Euro 2020* page (12,100,455 views), the second most viewed, just after the *Main Page* (554,030,839). There is a clear presence of articles that respond to general interests such as the *Bible* (11,048,609) or *Cleopatra* (9,516,340) pages. This may indicate that some topics raise general interest and may not be time-related.

The number of talks of Wikipedia articles is often used in conjunction with other variables in the construction of models for controversy detection (Jang et al., 2016). This suggests that this metric may be useful for detecting such controversial content in a simple way. Among the 20 pages with the highest number of talks stand out political figures, religion topics, and scientific controversies. The strong talk that takes place in some of them, as in *Donald Trump* (62,944), and the vandalism and presence of trolls, as in *Gamergate controversy* (27,185), have caused the editing of these pages to be restricted. In fact, there are some articles clearly related to controversial or sensitive issues, such as *Climate change* (40,837) and *Homeopathy* (25,898). In this regard, Wikipedia itself offers a page with a curated list of controversial articles (https://en.wikipedia.org/wiki/Wikipedia:List_of_controversial_issues), with 13 of the 20 pages listed as of 4 July 2021.

Finally, based on the volume of referenced publications, that is all materials with an associated identifier (DOI, ISBN, arXiv ID...), it is also possible to identify what are the Wikipedia pages that cite more scientific publications. However, in this case there are many research annuals and bibliographic pages present among the 20 articles, for example *2018 in* *paleontology* with 569 referenced publications. These lists have been eliminated to select the top 20 articles with encyclopedic content. In these articles there is a clear presence of scientific content, especially in medicine, such as *Feminizing hormone therapy* (329) and *Alzheimer's disease* (277). However, there are also articles related to history, such as *History of Lisbon* (313) or *World War II* (264). This may suggest that the metric of the number of publications cited can be used as a proxy to identify Wikipedia articles that are more scholarly oriented.

## 5. Discussion

In this study we describe how Wikipedia is a complex system, involving numerous actors and elements, and whose rules and governance depend on the community itself (Jemielniak, 2012). It is not only one of the first and clearest examples of Web 2.0 but also one of the few that remains among the most visited websites and has not deviated from its initial objective. Far from that, over the years it has gained the acceptance and trust of many of those who initially looked at it with skepticism.

We describe many similarities between scientific publications and Wikipedia pages. Both have different typologies of documents, structured content, evaluation of content and use of links and bibliographic references. There are also notable differences. While scientific publications may have limited access and a more specialized audiences, Wikipedia's content and scope is more open and target to more general audiences. The live nature of Wikipedia is probably its main distinctive feature when compared to scientific publications. Such live nature of Wikipedia articles must be considered when conducting informetric research on Wikipedia. To help in this endeavor, we propose an informetric-inspired conceptual framework, proposing different metrics that pay attention to the different analytical dimensions of Wikipedia, such as article characteristics, outreach, or citations to scientific publications among others. Some of these metrics have been already explored in the literature, such as page views (Mittermeier et al., 2019, 2021), but never in a comprehensive conceptual framework. The informetric-inspired conceptual framework presented here is expected to be useful for any Wikipedia study involving informetric, scientometric, bibliometric or webometric perspectives. Similarly, different Wikipedia data sources have been identified and described, finding in their differences in coverage, volume, access, or data processing crucial aspects for their selection.

Alongside the conceptual analytical framework proposed, a knowledge graph of the English edition of Wikipedia has been built and shared openly (doi:10.5281/zenodo.6346899). The data are gathered under a comprehensive dataset that follows a relational model and can be used by anyone interested in the study of this encyclopedia from an informetric point of view. It combines different data sources that allow on the one hand to characterize any Wikipedia page, while also allowing to establish relationships between each other (e.g., between two articles, an article and a category or an article and a linked website or a scientific publication referenced in it). Together with the metadata and relations of Wikipedia pages, the data of their bibliographic references are also incorporated, which come from the dataset shared by Singh et al. (Singh et al., 2020). It is precisely in Wikipedia's bibliographic reference data where greater efforts are needed so that they can be efficiently accessed through its official sources such as dumps or the API.

The case study provides a descriptive overview of Wikipedia articles, in its English edition, suggesting interesting valuable analytical possibilities and highlighting the relationships and usefulness of the metrics described. Our results suggest that the low correlations among most of the metrics point to the fact that the analytical dimensions measured through them are rather distinct. The potential analytical usefulness of some of the metrics has been highlighted. For example, the number of Wikipedia page views can be seen as a metric of social attention; the number of talks of Wikipedia pages can be seen as a proxy of controversial topics; and the number of scientific references in Wikipedia pages can help identify scholarly-related content. The use of the quality levels derived from WikiProjects has proved to be useful, showing clear differences between the different levels, but has also provided an overview of the Wikipedia articles.

Finally, it is important to also mention some of the limitations of this work. First, not all possible Wikipedia metrics and their relationships have been explored (e.g., the relationship between pages and users, or the number of users who follow the pages, the so-called watchers, or the number of editions in other languages of given article). The use of large amounts of data and some specific sources leads to a loss of consistency. For example, the Wikipedia dump process takes several days without blocking the edits during that time, so they are not really a snapshot. This loss of consistency also occurs when using different sources, especially when combining 2021 Wikipedia data with references from a third-party dataset published in 2020. The knowledge graph and the case study are based on the English Wikipedia, however, future research should study whether the same relationships found in this study also hold for other languages as well as the existing relationships between language editions.

## Funding

This work was funded by the Spanish Ministry of Science and Innovation with grant number PID2019-109127RB-I00/SRA/10.13039/501100011033. Wenceslao Arroyo-Machado received an FPU Grant (FPU18/05835) from the Spanish Ministry of Universities. Daniel Torres-Salinas received support under the Reincorporation Programme for Young Researchers of the University of Granada. Rodrigo Costas is partially funded by the South African DSI-NRF Centre of Excellence in Scientometrics and Science, Technology and Innovation Policy (SciSTIP).

# Appendix

## S1 Appendix. Wikipedia knowledge graph dataset construction.

In this section we describe the construction of a dataset combining data from different sources. In this case we have chosen to use the English edition of Wikipedia as it is the most extensive and the most documented, but the process described can be applied to a large extent to any edition.

## 1. Data retrieval

We have used different data sources to build our database, seeking to create a relational dataset that includes as complete as possible information about the pages, categories and references as well as the relationships established between them.

First, through the Wikipedia dumps[2], the *page*, *category*, *categorylinks*, *externallinks*, *pagelinks* and *page props* tables were downloaded, all of them available in SQL format. We also downloaded here the 27 files with the metadata of the history of revisions produced on Wikipedia pages (stub-meta-history files). All dumps are as of 2021-07-01. The SQL dumps have been processed by kwnlp_sql_parser Python package[3] (we performed some modifications to apply it to some dumps that are not covered), whereas XML dumps have been parsed by MediaWiki XML Processing Python package[4]. Then, the Analytics Datasets[5] were used to obtain page views for a period of 3 months (April 1, 2021 to June 30, 2021). These are tab-delimited plain text files.

Finally, we used the Singh, West, and Colavizza (2020) dataset: "*Wikipedia Citations: A comprehensive dataset of citations with identifiers extracted from English Wikipedia*". This dataset includes Wikipedia page references, which have been directly parsed from Wikipedia page content dumps published in May 2020, these files being in XML format. Thanks to this, each reference appears together with different attributes, such as authors, resource type, identifiers or URLs. Not only references to scientific papers are included, but also to all the contents included in that section, such as links to websites. Because this dataset has been parsed directly from the content of Wikipedia pages a referenced resource may appear in different ways on various Wikipedia pages, just like URLs. To overcome this problem, the Wikimedia Foundation has created the WikiCite initiative[6], which aims to create a bibliographic database of resources cited in Wikipedia to curate them.

---

[2] https://dumps.wikimedia.org/ (accessed on August 4, 2022)
[3] https://github.com/kensho-technologies/kwnlp-sql-parser (accessed on August 4, 2022)
[4] https://github.com/mediawiki-utilities/python-mwxml (accessed on August 4, 2022)
[5] https://dumps.wikimedia.org/other/analytics/ (accessed on August 4, 2022)

## 2. Data preprocessing

The constructed dataset with which a Wikipedia Knowledge Graph is established is composed of 9 different files in *tsv* format connected to each other, being Wikipedia pages the main one and around which the different relationships are established. Table 1 summarizes the different files of which the dataset is composed, and Fig 1 includes the relational model.

**Table 1.** Description of the files included in the Wikipedia Knowledge Graph dataset.

| File | Type | Description | Size |
|------|------|-------------|------|
| *page* | core | All Wikipedia pages metadata (id, title, views...) | 5.7 GB |
| *category* | base | Wikipedia categories metadata | 100 MB |
| *page_property* | base | Wikipedia pages properties defined in the page content. | 1.07 GB |
| *pub* | base | Scientific outputs referenced in Wikipedia article pages. | 153 MB |
| *url* | base | External websites linked from Wikipedia pages | 4.03 GB |
| *page_category* | intermediate | Categories linked in Wikipedia pages | 3.52 GB |
| *page_link* | intermediate | Links between Wikipedia pages | 9.36 GB |
| *page_pub* | intermediate | Wikipedia article pages citations to scientific outputs | 59 MB |
| *page_url* | intermediate | Links to external websites from Wikipedia pages | 1.23 GB |

**Fig 1.** Entity-Relationship Diagram (ERD) of the Wikipedia Knowledge Graph dataset.

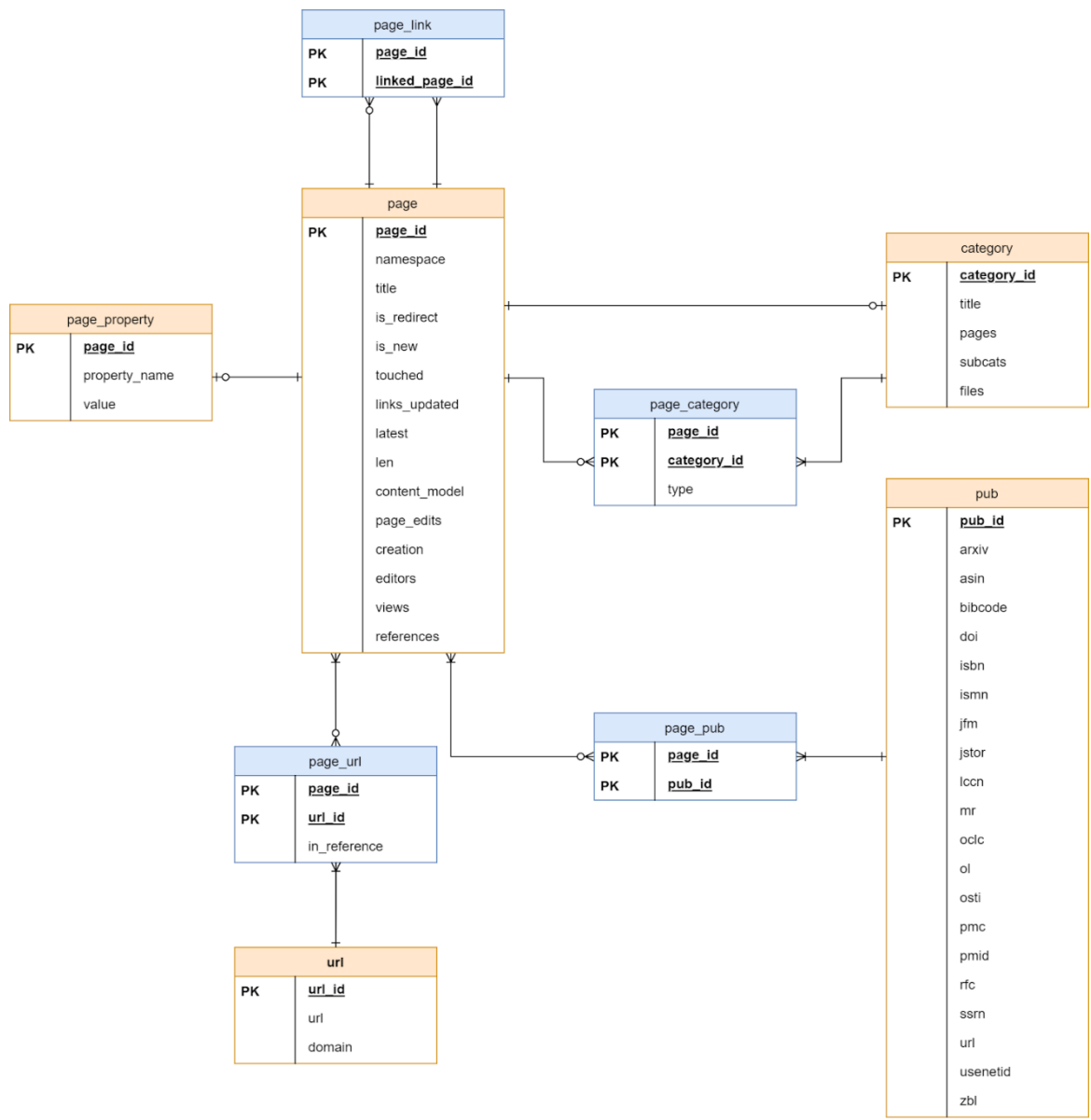

## 2.1. Pages

For the information on Wikipedia pages (*page.tsv — 53,710,529 x 15*), the SQL data dump has been used. It contains all the pages of all the namespaces, each of them uniquely identified by the *page_id*. Although there are many fields included in this table, these have been reduced to the title, namespace, restrictions, whether it is a redirection, whether the page is new (only has one revision or has been restored), timestamp of the last change and length.

In addition, the total number of views to the articles has been included using the information obtained for the three months. The total number of edits, number of unique editors and date of creation of a page have also been calculated by parsing the history of revisions files.

The page links are present in a file (*page_link.tsv — 566,536,991 x 2*) that has been preprocessed to not only reduce it to the main Wikipedia articles but to use the *page_id* instead of the page title to establish the links.

## 2.2. Categories

Because they are widely present on Wikipedia pages and not a biased subset and because they form a fundamental part of Wikipedia, categories have been used. In this case, no preprocessing has been applied and the two files from the SQL data dump have been used directly. Although the categories are Wikipedia pages (ns=14) and are included in the main table of pages, they are also treated independently from the rest of the pages. Thus, the first file (*category.tsv — 2,179,622 x 5*) provides for each category a unique identification number different from the page (*category_id*), its title, number of subcategories and files. The second file (*page_category.tsv — 165,501,704 x 3*) is the one that establishes the relationships between pages and categories. The only problem here is the identification of hidden categories, for which we have include the *page_props* file (*page_property.tsv — 28,967,070 x 3*) that provides additional information for Wikipedia pages, one of these fields being hiddencat, which indicates whether it is a hidden category or not.

## 2.3. References

Since the same bibliographic reference may appear on different Wikipedia pages but include different information, a standardization process has been carried out to harmonize this data. At all times minor changes have been made so as not to alter the editor's intention in including the reference.

First, the URLs to external pages have been reviewed. In this case we have two datasets, the Wikipedia dump and the dataset from Singh, West, and Colavizza (2020). The first one collects all the external URLs that appear on a Wikipedia page while the second one only those located in the references, so there is an overlap between the two. In general, common changes have been applied to all links in both datasets in parallel, eliminating the HTTP protocol part or the slashes at the end of the URL and parsing the URL, as well as fixing data errors such as spaces or line breaks. In cases where the URL was from an archiving service such as Wayback Machine or archive.today, it was shortened to the original URL. After that, the 50 most linked domains in both sets were selected—in the first dataset 27.70% of the links, while in the second dataset 20.02%—and specific errors in their URLs were corrected. For example, links to Google Books resources may include numerous optional parameters in their URL, such as the language of the interface and the exact page number, and after the cleanup all of them have been reduced to the identifier. Finally, both sets of links were combined into a single file (*urls.tsv — 51,923,982 x 3*), while the file that establishes the relationship between pages and external URLs (*page_url.tsv — 65,554,992 x 3*) indicates whether the URL appears in the references or not.

In the case of publications, no major changes were made. On the one hand, the identifiers of the publications were identified, using the 20 most common ones, and common errors in the DOI and ISBN, the two main ones, were corrected. Although an attempt was made to group the publications using all the unique identifiers, this option was discarded because if a chapter includes the DOI of the chapter and the ISBN of the book, the grouping would entail grouping all the chapters under the book. Instead, and to avoid this problem, we chose to directly identify each publication by the set of identifiers used in each reference. This resulted in a file containing the publications (*pub.tsv — 2,367,548 x 21*) and other with the connections between pages and referenced publications (*page_pub.tsv — 3,728,522 x 2*).

**S2 Table. Comparison of the evaluation process of Wikipedia pages and open peer review journals.**

| | Wikipedia | Open peer review |
|---|---|---|
| *Open identities* | The name of the wikipedian who edits and comments, or the IP address, is always visible | Authors and reviewers know each other's identity |
| *Open reports* | The discussion is fully accessible on the talk pages | The review reports are published together with the article |
| *Open participation* | Anyone can participate | The community can participate |
| *Open interaction* | Any wikipedian can discuss with another one directly through the talk pages | Discussion among the different people involved is allowed and encouraged |
| *Open pre-review manuscripts* | Pages can be published and edited without prior review or discussion of their content | Manuscripts are available prior to peer review |
| *Open final-version commenting* | Editing and discussion always remain open | Review or commenting on final publication |
| *Open platforms* | The whole process is done within Wikipedia | The review is provided by an entity other than the place of publication |

# S3 Appendix. Wikipedia data sources.

## 1. Wikimedia Dumps

This is a collection of files containing all the public data to date for some of the main database tables of each linguistic edition of Wikipedia, in addition to other Wikimedia Foundation project wikis. There are numerous dumps for each Wikipedia, the English one has almost 50, which include from textual content of the pages to page metadata or different types of relationships, as well as other types of entities, such as categories. For example, the *page* table includes page metadata, such as id, title, namespace or length in bytes, while the *pagelinks* table collects the internal links between pages.

However, Wikimedia Foundation warns of the limitations of these data dumps[7]. First of all, they are not backups, as they only contain public data. They do not provide a consistent state of the data, as due to the way the data is generated it is possible that modifications may occur during the process. It is not a snapshot of the database at a precise moment, but over a period of time the data is recovered and not in a structured way, so it is possible that inconsistencies are generated. Also, during all these years, broken data has been accumulating, which prevents these dumps from including a complete picture of Wikipedia.

Wikipedia dumps are generated once or twice a month and each one is available for 3 months, with some older dumps available at Internet Archive[8]. However, each time a dump is made it includes all the information available in Wikipedia, not the latest modifications. The only limitation is in the variations that the data structure may undergo, adding or deleting files or attributes. Regarding the licensing of this data, all original textual content is licensed under the Creative Commons Attribution-Share-Alike 3.0 License (CC BY-SA 3.0) and the GNU Free Documentation License (GFDL). Images and other files are available under different terms.

Although this data comes from the Wikimedia database, the dumps do not use a single format, with the different tables being in SQL or XML. This is due to a problem with Wikipedia's data structure and processing, which makes it much more efficient to provide table dumps as "text" in XML rather than SQL[9]. Also, through this way it is not possible to access all the data available in the database, being unavailable personal or sensitive information, such as information about users, as well as content that has been deleted[10].

---

[7] https://meta.wikimedia.org/wiki/Data_dumps/What_the_dumps_are_not (accessed on August 4, 2022)
[8] https://archive.org/details/wikimedia-mediatar (accessed on August 4, 2022)

The community itself offers good support for handling these files, as there are a wide variety of tools in different programming languages for decompressing multistream files, parsing XML and SQL files or processing them. In this sense, Mediawiki-utilities[11] is a collection of Python scripts that not only allow working with these data dumps, being able to parse both SQL and XML files, but also allow processing their content or querying APIs, among others. However, and as a function of Wikipedia, retrieving specific data, such as the number of edits to a page or the pages that link to it, involves costly data processing in this way.

## 2. MediaWiki and Wikimedia APIs

One of the main access points to the data is in the API, finding several for accessing data from any Wikipedia. Nevertheless, the main ones are the two MediaWiki APIs (MediaWiki Action API and MediaWiki REST API) and the Wikimedia REST API.

MediaWiki Action API allows not only to perform queries and retrieve entity metadata for each of the Wikimedia Foundation projects, not just Wikipedia, but also gives access to services such as authentication, content editing, creating a bot, or querying connected users. Although most of the public data queries do not require any kind of registration or password, options such as editing do require it. On the other hand, the MediaWiki REST API is also available for searching and querying pages, media files and page history. It has a more popular and simpler architecture, but does not offer as many options as the first one. Finally, the Wikimedia REST API gives access to content and metadata of Wikipedia pages and is more prepared for large volumes of requests. It also offers extra functionalities, such as obtaining pages related to a given one, generating a PDF of a page or specifically querying the talk pages. As in the previous ones, the endpoint (URL that serves as an entry point to the API services) is the Wikimedia project and the specific language edition from which data is to be retrieved. But,

---

in this case, it is possible to use wikimedia.org as an endpoint to access Wikipedia metrics such as page views or edits.

The main advantage this offers is not only access to atomic data without having to process a large volume of data, but also that the data consulted are up-to-date and can be consulted in real time. In addition, the content offered and its structure differ from that offered through data dumps. For example, while it is possible to obtain the number of watchers (users who have a page in their watchlist) for a page through the API, this is not available in the dumps, as it should be calculated through the watchlist table, which for privacy reasons does not appear publicly. This also happens with other metrics, such as the number of page edits, which can be calculated from the dumps, but requires effort in data processing.

In all three cases the documentation is well documented. The main problem with this tool is when large amounts of data need to be queried and retrieved. There is a limit to the number of requests, so in such situation retrieval may be less efficient than if a data dump is accessed directly. However, in some cases it is possible to request special permissions to raise this limit and there is even a Wikimedia Enterprise API.

## 3. Wiki Replicas

Another outstanding option for data recovery is Wiki Replicas, which are part of the services and products offered by Wikimedia Cloud Services (WMCS), previously known as Wikimedia Labs. Wiki Replicas are real-time and cleaned-up version copies of the WikiMedia databases with privacy-sensitive data removed. In addition, there are more tables included than those available in the dumps. Wiki Replicas are only accessible online and there are different tools and possibilities for this.

Toolforge is a Platform as a Service (PaaS) that offers a hosting and development environment in the cloud to work with WikiMedia Foundation projects. Among the functionalities offered is access to Wiki Replicas. Among the applications created under Toolforge highlights PAWS[12], a Jupyter Notebook environment with support for multiple programming languages that can also access Wiki Replicas, facilitating data analysis. On the other hand, Quarry[13] offers an interface to directly perform SQL queries to Wiki Replicas. In a similar way to what happens with the API, although these options open the door to a large amount of data, it is not possible to recover or work directly with large volumes, since cloud resources are limited for this purpose.

---

[12] https://wikitech.wikimedia.org/wiki/PAWS (accessed on August 4, 2022)

## 4. Specific data sources

The data sources discussed above are not only the main points of access to Wikipedia data but also cover data of numerous types. However, there is a wide range of possibilities both within and outside the Wikimedia ecosystem, in particular for the collection of more specific data, such as page views or activity logs.

Among the options provided by the platform itself is EventStreams, which offers data streams. With this it is possible to perform a completely live tracking, monitoring all the activity that takes place on Wikipedia. Analytic dumps and WikiStats are also available. The first is a collection of datasets containing quantitative metrics of Wikipedia activity, highlighting page views, disaggregated by time and day, devices and type of user (user, spider or automated), and the so-called clickstream that collects user clicks, although only for specific editions (English, Russian, German, Spanish, and Japanese). The second is a platform that offers statistical reports. This can be consulted through its web interface[14] or through the Wikimedia REST API.

Outside the Wikimedia environment there are other data sources. Of all of them, DBpedia stands out, since it offers a Wikipedia knowledge graph, establishing not only connections between the different elements that integrate the pages and are related to them but also providing semantics. As a result, they offer an extensive Open Linked Data dataset. In line with the statistical data, there is Xtools, which has a web interface and API and provides numerous statistical metadata of Wikipedia pages, such as views, edits, top editors or assessment. Finally, it is also possible to find datasets in repositories, highlighting those related to citations and bibliographic references (Halfaker et al., 2019; Singh et al., 2020) ,since this information has to be parsed from the pages.

---

[13] https://meta.wikimedia.org/wiki/Research:Quarry (accessed on August 4, 2022)
[14] https://wikitech.wikimedia.org/wiki/Analytics/Systems/Wikistats_2 (accessed on August 4, 2022)

**S4 Table. Top 20 English Wikipedia pages with the highest number of views (between 1 April and 30 June 2021), number of talks, and number of publications referenced.**

| | Title | Views* | Title | Talks** | Title | Publications referenced*** |
|---|---|---|---|---|---|---|
| 1 | UEFA Euro 2020 | 12,100,455 | Donald Trump | 62,944 | Soil | 439 |
| 2 | Deaths in 2021 | 11,540,360 | Barack Obama | 46,623 | Evolution | 362 |
| 3 | Bible | 11,048,609 | Climate change | 40,837 | Neanderthal | 353 |
| 4 | Prince Philip, Duke of Edinburgh | 10,860,553 | Intelligent design | 32,564 | Feminizing hormone therapy | 329 |
| 5 | Microsoft Office | 10,270,633 | United States | 31,296 | History of Lisbon | 313 |
| 6 | Elizabeth II | 9,900,275 | Jesus | 30,617 | Bioelectricity | 294 |
| 7 | Cleopatra | 9,516,340 | Sarah Palin | 28,514 | Murine respirovirus | 293 |
| 8 | Google logo | 9,040,065 | Gamergate controversy | 27,185 | Soviet dissidents | 290 |
| 9 | XXXX | 7,008,866 | Homeopathy | 25,898 | Induced stem cells | 289 |
| 10 | Elon Musk | 6,667,656 | Race and intelligence | 25,565 | Bicalutamide | 281 |
| 11 | Mare of Easttown | 5,995,513 | Gaza War (2008–2009) | 25,179 | Alzheimer's disease | 277 |
| 12 | Mortal Kombat (2021 film) | 5,735,431 | September 11 attacks | 24,272 | Schizophrenia | 270 |
| 13 | Godzilla vs. Kong | 5,566,200 | Muhammad | 23,872 | Eating disorder | 267 |
| 14 | Critical race theory | 5,428,240 | India | 23,795 | Castellania (Valletta) | 267 |
| 15 | Charles Sobhraj | 5,321,586 | Evolution | 23,487 | Pharmacokinetics of estradiol | 267 |
| 16 | Skathi (moon) | 5,301,961 | George W. Bush | 23,097 | African humid period | 267 |
| 17 | Cristiano Ronaldo | 4,819,147 | Circumcision | 22,772 | Haplogroup T-M184 | 265 |
| 18 | The Falcon and the Winter Soldier | 4,736,901 | Climatic Research Unit email controversy | 22,686 | World War II | 264 |
| 19 | Army of the Dead | 4,494,209 | Cities and towns during the Syrian civil war | 22,317 | Canada | 263 |
| 20 | Charles, Prince of Wales | 4,453,451 | Prem Rawat | 22,271 | Temporal envelope and fine structure | 261 |

\* Main Page has been omitted, which is the first one with 554,030,839 views.
\*\* Main Page has been omitted, which is the first one with 151,174 talks.
\*\*\* Research annuals and bibliographic pages have been omitted.